# Dynamic Bragg microcavities in collisions of unipolar light pulses of unusual shape in two- and three-level medium

Rostislav Arkhipov[1], Mikhail Arkhipov[1], Nikolay Rosanov[1]

[1]Ioffe Institute, St. Petersburg, Russia

e-mail: arkhipovrostislav@gmail.com, mikhail.v.arkhipov@gmail.com, nnrosanov@mail.ru

Unipolar light pulses with a non-zero electric area due to the unidirectional action on charged particles can be used for the ultrafast control of the properties of quantum systems. To control atomic properties in an efficient way, it is necessary to vary the temporal shape of the pulses used. This has led to the problem of obtaining pulses of an unusual shape, such as a rectangular one. A number of new phenomena, not possible with conventional multi-cycle pulses, were discovered by analyzing the interaction of such unipolar pulses with matter. These include the formation of dynamic microcavities at each resonant transition of a multilevel medium when such pulses collide with the medium. In this work, we compare the behavior of dynamic microcavities in a two-level and a three-level medium when unipolar pulses of unusual shape (rectangular) are collided with the medium. We do this on the basis of the numerical solution of the system for the density matrix of the medium and the wave equation for the electric field. Medium parameters correspond to atomic hydrogen. It is shown that for rectangular pulses in a three-level medium, the dynamics of the cavities can be very different from the two-level model, as opposed to pulses of other shapes (e.g. Gaussian shape). When the third level of the medium is taken into account, the self-induced transparency-like regime disappears. Differences in the dynamics of resonators in a three-level medium are revealed when the pulses behave like $2\pi$ pulses of self-induced transparency.

# Introduction

Advances in reducing the duration of electromagnetic pulses in recent years have led to a great deal of work on obtaining electromagnetic pulses of attosecond duration [Midorikawa 2022, Kim et al. 2023, Sevrino et al. 2024]. Such pulses make it possible to study the dynamics of electrons in matter in great detail [Huillie 2024, Krausz 2024, Agostini 2024]. Underlining the importance of research in this direction, the Nobel Prize in Physics was awarded for research in this field [URL]. Further reduction of the pulse duration will inevitably lead to obtaining unipolar (or half-wave) pulses, consisting of one half-wave of the field. These are obtained when all half-waves are removed from an ordinary multi-cycle pulse and one is left. The electrical area, defined as the integral of the electric field strength $\boldsymbol{E}(\boldsymbol{r},t)$ over time $t$ at a given point in space $\boldsymbol{r}$ [Jackson 1962, Bessonov 1981, Rosanov 2009, Rosanov et al. 2018], is an important physical quantity that applies only to half-wave pulses:

$$\boldsymbol{S}_E = \int \boldsymbol{E}(\boldsymbol{r},t)dt. \qquad (1)$$

Unipolar pulses were apparently first mentioned in Jackson's book in the problem of the calculation of the electric field strength of uniformly moving charges [Jackson 1962], where the value (1) is called the "time integral of the fields" (Ch.11, p.554 in [Jackson 1962]). At about the same time, V.L. Ginzburg and his colleagues showed that the magnetic bremsstrahlung radiation of a charge moving in a circle (synchrotron radiation) is a sequence of half-cycle pulses [Ginzburg et al. 1966]. Bessonov's work [Bessonov 1981] showed the possibility of obtaining unipolar pulses with accelerated charge motion. Such pulses are called "strange waves" in this paper. In the optics of extremely short pulses, however, this term does not seem to be very successful. The term "electric pulse area" was introduced in [Rosanov 2009]. In this paper, the rule for the conservation of the pulse area in the one-dimensional case ("Rosanov's rule", see also review [Rosanov et al. 2018]) was established.

In recent years, interest in the generation of unipolar pulses [Wu et al. 202, Hassan et al. 2016, Xu et al. 2018, Bogatskaya et al. 2022, Ilyakov et al. 2022, Arkhipov et al., 2023, Glazov et al., 2024, Rosanov et al. 2024, Arkhipov et al, 2022, Gorelov et al. 2025] and the characteristics of their interaction with matter has grown rapidly [Rosanov et al., 2021, Pakhomov et al.,2024, Arkhipov et al. 2024, Alexandrov et al., 2024 ]. The nonlinear optics of unipolar pulses is developing rapidly. It has become a new direction in modern physics. The results of the most recent studies in this field are summarized in the reviews [Arkhipov et al., 2020, Arkhipov et al., 2023, Rosanov et al., 2023, Rosanov et al., 2023, Rosanov et al., 2024] and in a chapter of the monograph [Rosanov et al. 2023]. Unipolar pulses can rapidly deliver a unidirectional pulse to a charged particle. This fact makes them a promising tool for the ultrafast excitation of quantum systems and the study of the dynamics of electrons at times shorter than the characteristic time associated with the energy of the ground state (the period of one electron revolution along the Bohr orbit in an atom) [Rosanov et al., 2021, Pakhomov et al.,2024, Arkhipov et al. 2024]. Recent experimental results show the possibility of obtaining THz unipolar pulses [Arkhipov et al. 2022, Gorelov et al. 2025].

An additional problem with the use of unipolar pulses for these applications is the control of their time shape. Recently, work has appeared demonstrating the possibility of obtaining unipolar pulses of unusual shape, e.g. rectangular. (Such pulses are known in the radio range; in optics, the problem of obtaining such control pulses has not yet been considered) [Ilyakov et al. 2022, Kuratov et al. 2022, Sazonov et al., 2018, Pakhomov et al. 2022, Arkhipov et al. 2023, Pakhomov et al. 2023]. Such pulses are interesting for exciting atoms [Arkhipov et al. 2022], superconducting qubits [Bastrakova et al. 2019, Bastrakova et al. 2020]. Their interaction with matter, however, has been the subject of little research to date. Studying the interaction of unipolar pulses with a medium has led to the prediction and further active study of a number of new phenomena that seem impossible using conventional multi-cycle pulses [Arkhipov et al., 2020, Arkhipov et al., 2023, Rosanov et al., 2023, Rosanov et al., 2023, Rosanov et al., 2024].

One such phenomenon is the formation of dynamic microcavities (microresonators, DM) upon collision of half-cycle pulses in a medium [Arkhipov et al. 2022, Diachkova et al. 2023, Diachkova et al. 2024, Arkhipov et al., 2024, Arkhipov et al., 2024, Arkhipov et al. 2024], see also review [Arkhipov et al. 2023, 2024]. When such pulses collide in a medium, the population difference may be constant in the region where they overlap and suddenly change outside this region. Or, outside the pulse overlap region, a Bragg grating of atomic populations can form on either side of the overlap region. This means that a DM is created in the medium. The parameters of this DM can be easily controlled when the pulses collide again in the medium.

Apparently, the emergence of such structures was first suggested in [Arkhipov et al. 2022], which studied the collision of rectangular femtosecond pulses in a two-level medium with a small amplitude. In [Diachkova et al. 2023], the formation of DM was studied during the collision of $2\pi$-like rectangular self-induced transparency (SIT) pulses, also in a two-level medium. In this work, it was shown that the population difference has an almost constant value in the pulse overlap region and another constant value outside. During subsequent collisions between pulses, the value of the population difference in the overlap region changed. This indicates the possibility of controlling the properties of DM.

The behavior of DM during collision of Gaussian shaped unipolar attosecond pulses in a two-level medium in weak and strong fields, when the SIT regime is realized, was considered in [Arkhipov et al. 2024]. An analytical approach based on an approximate solution of the Schrödinger equation is also proposed in [Arkhipov et al. 2024]. This approach shows the formation of the DM at each resonant transition of a multilevel medium. However, this approach is only valid for a small amplitude of the field of the excitation pulses. The numerical calculations carried out in [48-

Arkhipov et al. 2024] for a three-level medium have shown a qualitative agreement between the shape of the DM in the SIT regime and the case of a two-level medium analyzed in [46- Arkhipov et al. 2024].

The characteristics of the DM in the collisions of rectangular and triangular pulses (not in the SIT regime) in a three-level medium with model parameters and a low particle density have been considered in [45-Arkhipov et al. 2024]. However, a direct comparison of DM behavior with the results of DM calculations in a two-level model of the medium, especially when realizing the SIT regime, was not performed in this work. The question of the applicability of the two-level approximation naturally arises when analyzing the dynamics of the DM in strong fields. In [49,50- Arkhipov et al. 2024, Arkhipov et al. 2024] a comparison was made between the dynamics of population gratings for the case of non-overlapping pulses in a two-level medium and taking into account the third level in the medium. Both similarities and differences have been shown.

Some qualitative remarks of a physical nature on the conservation of the DM in a multilevel medium have been discussed in [49,50- Arkhipov et al. 2024, Arkhipov et al. 2024] and in the review [47- Arkhipov et al, 2024, 2023]. The conservation of DM and population gratings in a multilevel medium, which was originally predicted in a two-level medium, follows from the physical mechanism of their formation. The coherent interaction of pulses with the medium is the basis of this mechanism. However, the shape and parameters of the DM can change significantly when additional levels are taken into account. Therefore, a direct comparison of their behavior in a two- and multilevel medium is necessary for a complete understanding of DM dynamics in real multilevel media.

In this paper, based on the numerical solution of the system of equations for the density matrix of a two- and three-level medium together with the wave equation for the electric field strength, a comparison of the DM dynamics in these cases will be made when a sequence of rectangular pulses collides in the medium. The parameters of the medium (dipole moments and resonant transition frequencies) are considered as in the case of the hydrogen atom. This illustrates the occurrence of these effects in real media.

It is shown that for rectangular pulses, as opposed to Gaussian pulses [Arkhipov et al. 2024, Arkhipov et al. 2024- Refs.46,47], the dynamics of the structures studied can change significantly when the third level is included. However, when the excitation amplitude is small (not in the SIT-like regime), including the additional level leads to an insignificant change in the shape of the induced DMs. On the other hand, in the SIT regime, significant differences in the shape of the DM in a three-level medium are found compared to the case of a two-level medium. It is also shown that the SIT-like propagation of rectangular pulses disappears when the third energy level is added. This is in contrast to the case of Gaussian and hyperbolic secant pulses studied by the authors previously [Arkhipov et al. 2024, Arkhipov et al. 2024- Refs.51,52].

## Model and system under study

A series of numerical calculations have been carried out. The medium has been modelled in two and three level approximations. The system of equations for the density matrix of a two-level medium, together with the wave equation, is of the form given in [Yariv 1975]:

$$\frac{\partial \rho_{12}(z,t)}{\partial t} = -\frac{\rho_{12}(z,t)}{T_2} + i\omega_0 \rho_{12}(z,t) - \frac{i}{\hbar} d_{12} E(z,t) n(z,t), \qquad (2)$$

$$\frac{\partial n(z,t)}{\partial t} = -\frac{n(z,t) - n_0(z)}{T_1} + \frac{4}{\hbar} d_{12} E(z,t) \text{Im}\rho_{12}(z,t), \qquad (3)$$

$$P(z,t) = 2N_0 d_{12} \text{Re}\rho_{12}(z,t), \qquad (4)$$

$$\frac{\partial^2 E(z,t)}{\partial z^2} - \frac{1}{c^2}\frac{\partial^2 E(z,t)}{\partial t^2} = \frac{4\pi}{c^2}\frac{\partial^2 P(z,t)}{\partial t^2}. \qquad (5)$$

This system of equations (2)-(5) contains the following model parameters: $E$ is the electric field strength, $P$ is the polarization of the medium, $t$ is the time, $z$ is the longitudinal coordinate, $d_{12}$ is the transition dipole moment, $c$ is the speed of light in a vacuum, $\omega_0$ is the transition frequency, $N_0$ is the concentration of two-level particles, $\hbar$ is the reduced Planck constant, $T_1$ is the relaxation time of the population difference, $T_2$ is the relaxation time of the polarization of the medium, $n_0$ is the population difference of the medium in the absence of an electric field ($n_0 = 1$ for an absorbing medium), $\rho_{12}$ is off-diagonal element of the density matrix of a two-level medium, $n = \rho_{11} - \rho_{22}$ is the population difference (inversion) of a two-level medium. It should be added that one-dimensionality is justified in the case of a coaxial waveguide [Rosanov et al. 2023].

In the case of a three-level medium, the system of equations of the model has the form [Yariv 1975]:

$$\frac{\partial}{\partial t}\rho_{21} = -\rho_{21}/T_{21} - i\omega_{12}\rho_{21} - i\frac{d_{12}}{\hbar}E(\rho_{22} - \rho_{11}) - i\frac{d_{13}}{\hbar}E\rho_{23} + i\frac{d_{23}}{\hbar}E\rho_{31}, \qquad (6)$$

$$\frac{\partial}{\partial t}\rho_{32} = -\rho_{32}/T_{32} - i\omega_{32}\rho_{32} - i\frac{d_{23}}{\hbar}E(\rho_{33} - \rho_{22}) - i\frac{d_{12}}{\hbar}E\rho_{31} + i\frac{d_{13}}{\hbar}E\rho_{21}, \qquad (7)$$

$$\frac{\partial}{\partial t}\rho_{31} = -\rho_{31}/T_{31} - i\omega_{31}\rho_{31} - i\frac{d_{13}}{\hbar}E(\rho_{33} - \rho_{11}) - i\frac{d_{12}}{\hbar}E\rho_{32} + i\frac{d_{23}}{\hbar}E\rho_{21}, \qquad (8)$$

$$\frac{\partial}{\partial t}\rho_{11} = \frac{\rho_{22}}{T_{22}} + \frac{\rho_{33}}{T_{33}} + i\frac{d_{12}}{\hbar}E(\rho_{21} - \rho_{21}^*) - i\frac{d_{13}}{\hbar}E(\rho_{13} - \rho_{13}^*), \qquad (9)$$

$$\frac{\partial}{\partial t}\rho_{22} = -\rho_{22}/T_{22} - i\frac{d_{12}}{\hbar}E(\rho_{21} - \rho_{21}^*) - i\frac{d_{23}}{\hbar}E(\rho_{23} - \rho_{23}^*), \qquad (10)$$

$$\frac{\partial}{\partial t}\rho_{33} = -\frac{\rho_{33}}{T_{33}} + i\frac{d_{13}}{\hbar}E(\rho_{13} - \rho_{13}^*) + i\frac{d_{23}}{\hbar}E(\rho_{23} - \rho_{23}^*), \qquad (11)$$

$$P(z,t) = 2N_0 d_{12}\mathrm{Re}\rho_{12}(z,t) + 2N_0 d_{13}\mathrm{Re}\rho_{13}(z,t) + 2N_0 d_{23}\mathrm{Re}\rho_{23}(z,t), \qquad (12)$$

$$\frac{\partial^2 E(z,t)}{\partial z^2} - \frac{1}{c^2}\frac{\partial^2 E(z,t)}{\partial t^2} = \frac{4\pi}{c^2}\frac{\partial^2 P(z,t)}{\partial t^2}. \qquad (13)$$

In this system of equations, the variables $\rho_{11}, \rho_{22}, \rho_{33}$ are the populations of the 1st, 2nd and 3rd states of the medium, respectively; $\rho_{21}, \rho_{32}, \rho_{31}$ are the off-diagonal elements of the density matrix which determine the dynamics of the polarisation of the medium; $d_{12}, d_{13}, d_{23}$ are the dipole moments of the transitions; $\omega_{12}, \omega_{32}, \omega_{31}$ are the frequencies of the resonance transitions; $T_{ik}$ are the relaxation times. Polarization relaxation times can reach values of tens and hundreds of nanoseconds in gaseous media and solids cooled to helium temperatures [Bayer et al. 2002]. These values are even greater, ranging from seconds to hours, in crystals of rare earth ions [Rabbit et al. 1988]. These values are much larger than the pulse durations and the intervals between them. Therefore, for times shorter than the relaxation time, relaxation can be neglected. The values of the medium parameters (dipole moments and resonance transition frequencies) are taken from the book [Frisch 1963]. They correspond to atomic hydrogen.

The model assumes that the medium is placed in a resonator with ideal mirrors, and that the length of the resonator is $L = 12\lambda_0$. And the medium itself was located in the region between points with coordinates $z_1 = 2\lambda_0$ and $z_2 = 10\lambda_0$. The pulses reflected from the ideal mirrors returned to the medium where they collided at point, $z_{col} = 6\lambda_0$. Such a scheme with ideal mirrors is chosen only for the convenience of the numerical calculations. It allows one to easily generate a sequence of counter-pulses with the required number of collisions in the medium. Such a scheme is obviously idealized from a practical point of view. In practice, a specially modernized optical scheme with mirrors and a source of half-cycle pulses must be used to generate a sequence of counter-pulses with the required number of collisions.

The density matrix equation system was solved by the 4th order Runge-Kutta method, the wave equation by the finite difference method with a step of $\Delta z = \lambda_0/200$. A hyper-Gaussian function was used to model the electric field of rectangular pulses. The impulses entering the medium from its left and right edges had the shape at the boundaries of the medium:

$$E(z=0,t) = E_0 e^{-\frac{(t-2.5\tau)^{20}}{\tau^{20}}}, \tag{14}$$

$$E(z=L,t) = E_0 e^{-\frac{(t-2.5\tau)^{20}}{\tau^{20}}}. \tag{15}$$

Such a scheme allowed to create a sequence of pulses in the numerical simulations. It allows the control of the DM after each collision of pulses.

Table. Parameters used in numerical calculations

| Frequency (wavelength $\lambda_0$) of the transition 1→2 | $\omega_{12} = 1.55 \cdot 10^{16}$ rad/s ($\lambda_{12} = \lambda_0 = 121.6$ nm) |
|---|---|
| Transition dipole moment 1→2 | $d_{12} = 3.27$ D |
| Frequency (wavelength) of the transition 1→3 | $\omega_{13} = 1.84 \cdot 10^{16}$ rad/s ($\lambda_{13} = 102.6$ nm) |
| Transition dipole moment 1→3 | $d_{13} = 1.31$ D |
| Frequency (wavelength) of the transition 2→3 | $\omega_{23} = 2.87 \cdot 10^{15}$ rad/s ($\lambda_{23} = 656.6$ nm) |
| Transition dipole moment 2→3 | $d_{23} = 12.6$ D |
| Atomic concentration | $N_0 = 10^{20}$ cm$^{-3}$ |
| Electric field amplitude | $E_0 = 2 \cdot 10^6$ ESU |

The duration and amplitude of pulses (14) and (15) were chosen based on the following considerations. In [57], an analytical solution of equations (2)-(3) was found (ignoring the relaxation times of the medium), describing the behavior of the polarization $P(t)$ and the population difference $n(t)$ under the action of rectangular pulses. It has the form for the duration of the pulse, at $0 \leq t \leq \tau$:

$$n(t) = \frac{4\Omega_R^2 \cos(\Omega t) + \omega_0^2}{\omega_0^2 + 4\Omega_R^2}, \quad P(t) = \frac{2\omega_0 d_{12}^2 E_0}{\hbar(\omega_0^2 + 4\Omega_R^2)}[1 - \cos(\Omega t)]. \tag{16}$$

Here frequency $\Omega \equiv \sqrt{\omega_0^2 + 4\Omega_R^2}$, whre $\Omega_R = d_{12}E_0/\hbar$ is the Rabi frequency.

It shows that in a two-level medium, the polarization and population difference oscillate not at the Rabi frequency, as in the case of a resonant multi-cycle pulse, but at the frequency $\Omega \equiv \sqrt{\omega_0^2 + 4\Omega_R^2}$, which depends on the Rabi frequency and the transition frequency. After the pulse has ended, the population difference remains constant, and the polarization oscillates at the transition frequency, which last for the relaxation time. It is clear from (16) that by selecting the pulse duration and its amplitude, it is possible to obtain the desired value of the inversion n after the pulse has passed. For example, the medium will return to the ground state with the inversion $n = 1$ (SIT regime [Allen et al. 1975, McCall et al. 1969]) if the pulse duration satisfies the condition:

$$\tau = \frac{\pi m}{\Omega} = \frac{\pi m}{\sqrt{\omega_0^2 + 4\Omega_R^2}}, \qquad (17)$$

where m=2,3,… . If $m = 1$, which corresponds to half the period of the frequency Ω, then according to (16) after the end of the pulse $n(t) = \frac{\omega_0^2 - 4\Omega_R^2}{\omega_0^2 + 4\Omega_R^2}$. This value can obviously also be made close to 1 at $\Omega_R \ll \omega_0$, which allows implementing the SIT regime mode at $m = 1$. For convenience, in the numerical calculations below, the value of the parameter τ was calculated using formula (17). In the first series of numerical calculations, in the weak field mode, when the SIT was not achieved, the value of m was taken to be less than 1. Then, the dynamics of the DM in the SIT mode was analyzed, when $m = 1$.

**Comparison of the dynamics of the DM in a two-level and in a three-level medium during the collision of rectangular pulses of small amplitude**

In the weak-field regime (beyond SIT), when $m = 0.6$, a numerical solution of the system of equations (2)-(5) was performed. In this case, the pulse amplitude $E_0 = 2 \cdot 10^6$ ESU and the parameter $\tau = 95$ as were taken as follows. The table shows the remaining parameters. To obtain a high value of Q-factor of the induced DM, large concentrations of atoms, $N_0 \cong 10^{20}$ cm$^{-3}$ are required [46- Arkhipov et al. 2024]. They are achievable through the placement of atomic hydrogen in the matrix [Ahokas et al. 2009, Järvinen et al. 2011-60,61].

Figures 1 and 2 show a typical spatio-temporal dynamics of polarization and population difference, similar to that considered previously in a two-level medium [Arkhipov et al. 2022]. It is calculated as a result of three collisions of pulses that occurred at the moments $t_{c1} = 2.7$ fs, $t_{c2} = 7.6$ fs, $t_{c3} = 12.6$ fs. It can be seen that in the region of pulse overlap near the point $z_{col} = 6\lambda_0$, after the first collision between the pulses, the population difference has an almost constant value. To the left and right of this point, Bragg gratings of the population difference are formed. This indicates the formation of a "quasi-resonator" in the medium.

Each subsequent pulse collision alters the shape of these DMs by changing the medium's polarization oscillations. The mechanism of formation of these gratings is known. It is based on the coherent interaction of the pulses with the medium [46,49,62-63- Arkhipov et al. 2024, Arkhipov et al. 2024, Arkhipov et al. 2016, Arkhipov et al. 2021]. After the pulses pass through the medium, polarization waves are formed in it (travelling or stationary waves of complex shape). These waves exist for a time $T_2$. The interaction of subsequent pulses with these medium coherence oscillations causes gratings to appear at each resonant transition of the multilevel medium [Arkhipov et al. 2024- Ref.46].

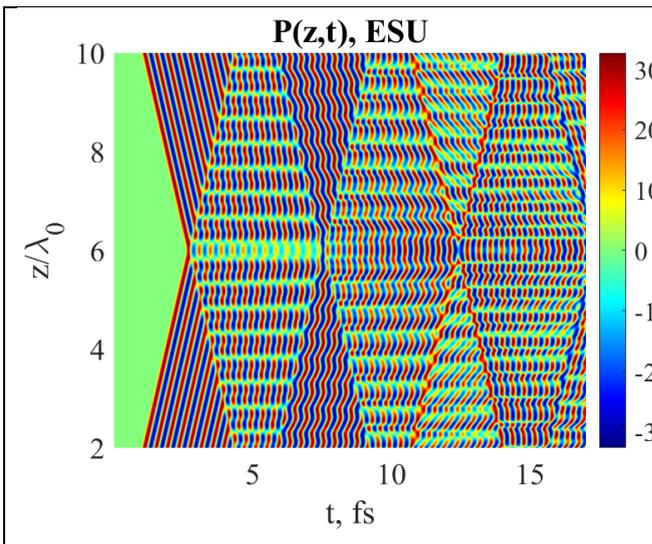

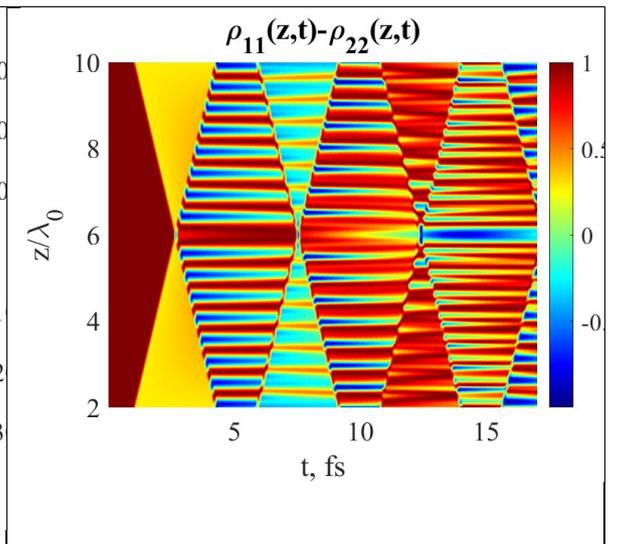

| Fig. 1. Spatio-temporal dynamics of polarization of a two-level medium $P(z,t)$. | Fig. 2. Spatio-temporal dynamics of the populations difference of a two-level medium $n(z,t) = \rho_{11} - \rho_{22}$. |

Polarization oscillations (medium coherence, non-diagonal elements of the density matrix) occur at other transitions of the medium when the additional level of the medium is taken into account. In our case of an atomic medium, the third level is close to the second, so the shape of the induced DM will not change much when the third level is taken into account. The behavior of polarization and population difference in a three-level medium obtained by numerically solving Eqs. (6)-(13) is shown in Fig. 3-7. It is obvious that after the first pulse collision, the shape of the resonator does not change much when the additional level is taken into consideration. However, due to the more complex nature of the polarization oscillations of the medium in a three-level medium, the shape of the resonator changes after subsequent collisions.

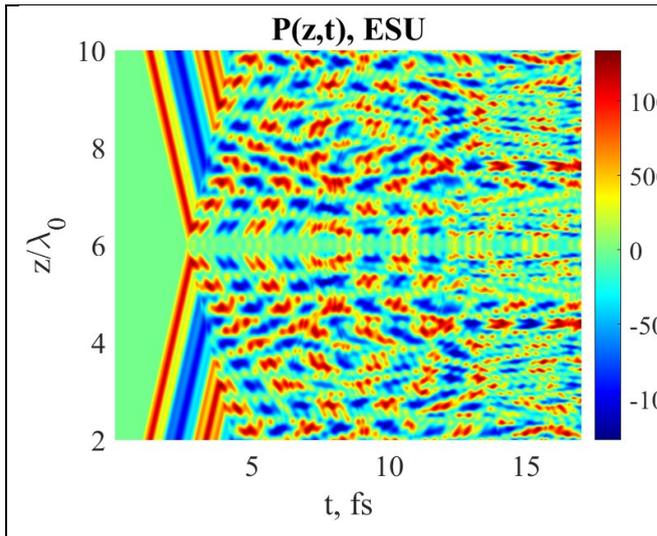

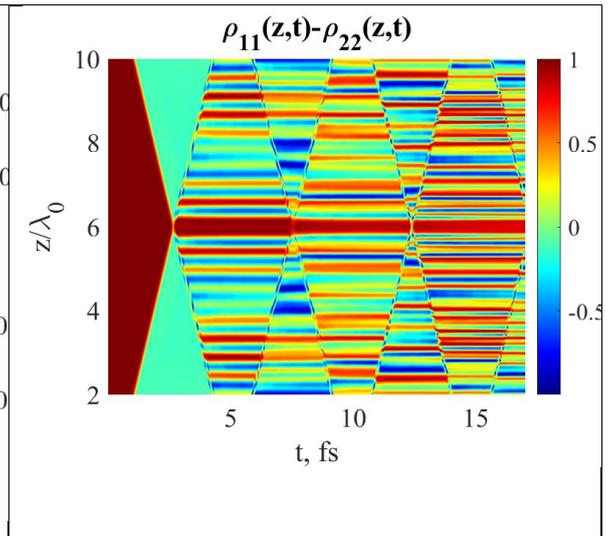

Fig. 3. Spatio-temporal dynamics of polarization of a three-level medium $P(z,t)$.

Fig. 4. Spatio-temporal dynamics of population differences $\rho_{11} - \rho_{22}$ of a three-level medium.

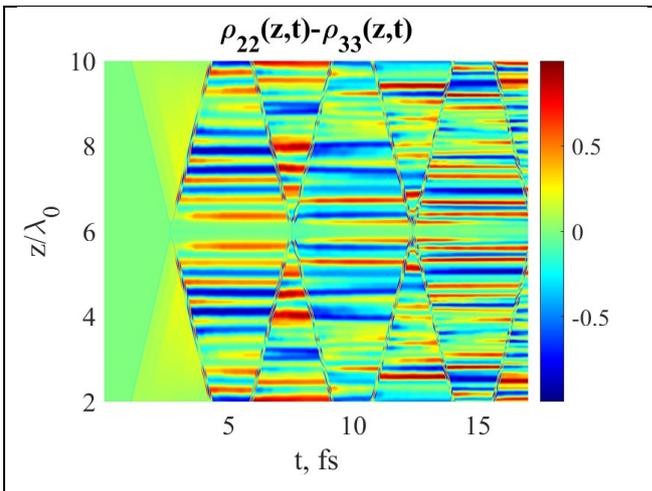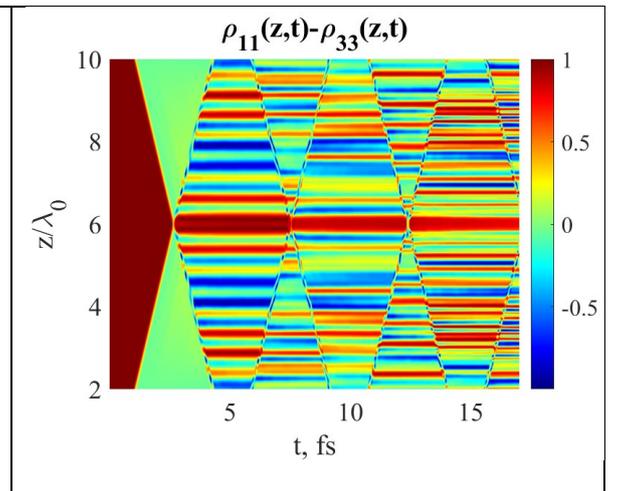

| **Fig. 5.** Spatio-temporal dynamics of population differences $\rho_{22} - \rho_{33}$ of a three-level medium. | **Fig.6.** Spatio-temporal dynamics of population differences $\rho_{11} - \rho_{33}$ of a three-level medium. |

## Comparison of DM dynamics in a two- and three-level medium upon collision of $2\pi$-like rectangular pulses of self-induced transparency

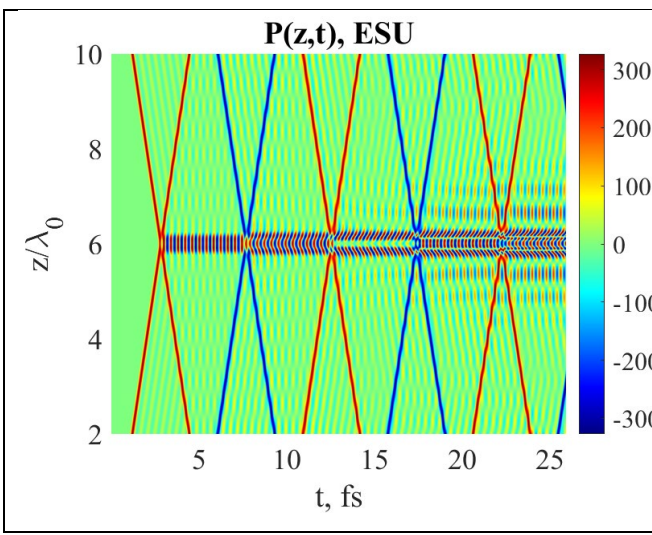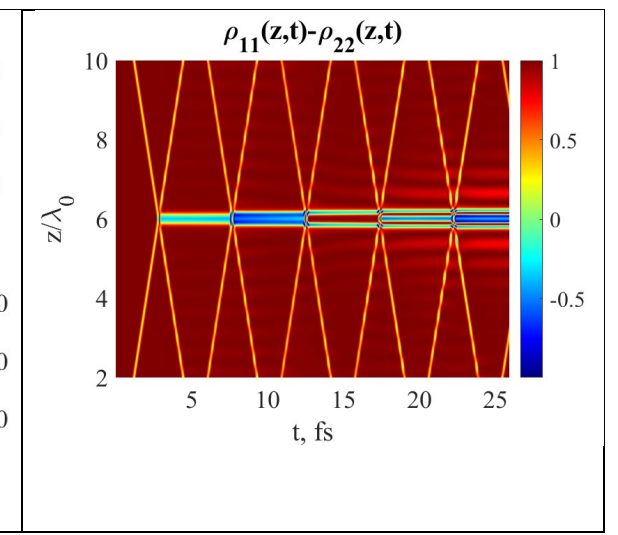

| **Fig.7.** Spatio-temporal dynamics of polarization of a two-level medium $P(z,t)$. $P(z,t)$. | **Fig.8.** Spatio-temporal dynamics of the populations difference of a two-level medium $n(z,t) = \rho_{11} - \rho_{22}$. |
|---|---|

The dynamics of the DM in a multi-level medium becomes more complex when the exciting pulses act like the $2\pi$ SIT-like pulses of the McCall and Hahn [McCall et al. 1969]: the first half of the pulse excites the medium, and the second half returns it to the ground state. In a two-level medium, as mentioned above, this can be realized when $m = 1$. For such parameters, $\tau = 158$ as is used. Figures 8-9 show the results of numerical calculations illustrating the behavior of polarization and population difference in a two-level medium. They have been obtained as a result of 5 collisions between the pulses. In the case of a three-level medium, this dynamic is shown in Fig. 9-12 with the same parameters as in the case of a two-level medium.

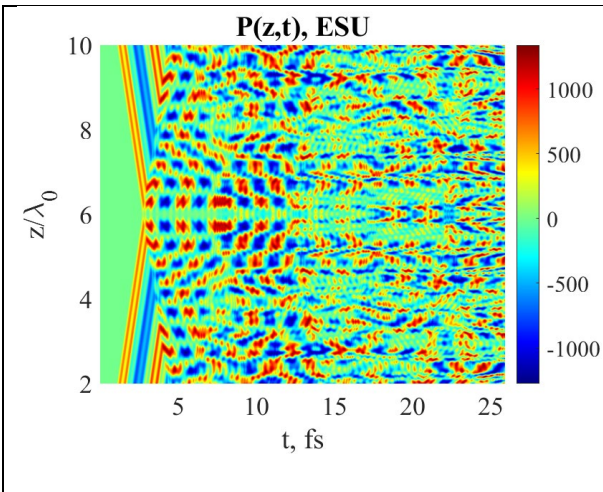 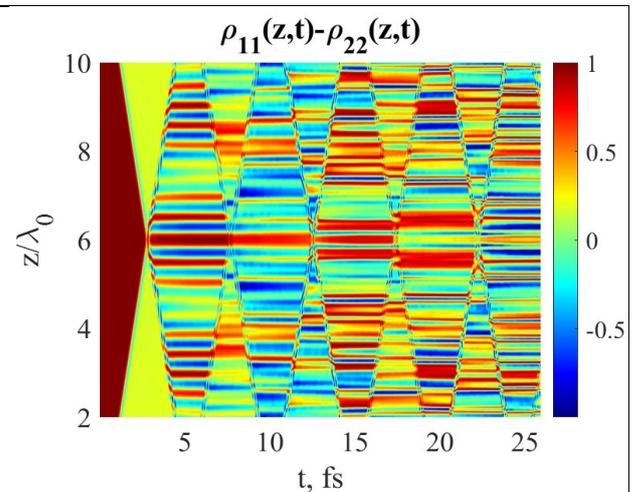

| **Fig. 9.** Spatio-temporal dynamics of polarization of a three-level medium $P(z,t)$. | **Fig.10.** Spatio-temporal dynamics of population differences $\rho_{11} - \rho_{22}$ of a three-level medium. |
|---|---|

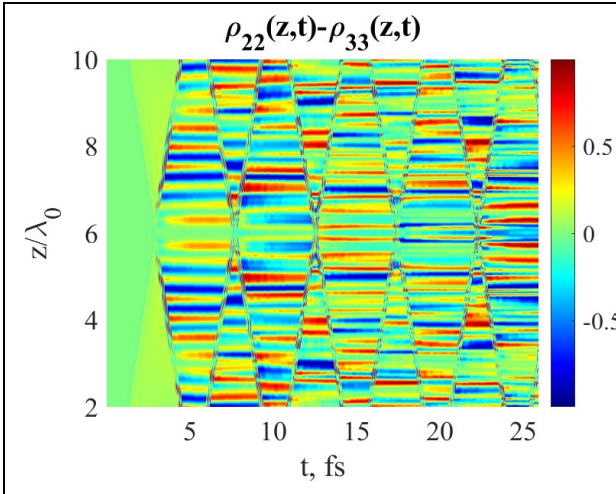 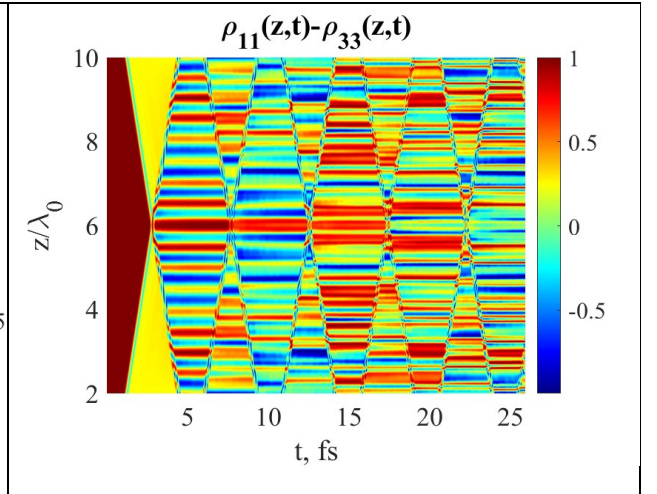

**Fig. 11.** Spatio-temporal dynamics of population differences $\rho_{22} - \rho_{33}$ of a three-level medium.

**Fig. 12.** Spatio-temporal dynamics of population differences $\rho_{11} - \rho_{33}$ of a three-level medium.

The behavior of the population difference is similar to that previously reported in [42, 44- Diachkova et al. 2023, Arkhipov et al. 2024] for a two-level medium. In the region of pulse overlap, the population difference has an almost constant value. This value differs after each pulse collision. It abruptly changes to another constant value outside this region. The result is a DM whose parameters are controlled by an increase in the number of collisions. In this case, to the left and right of the pulse overlap region, no Bragg population gratings were formed.

The behavior of the DM changes significantly upon transition to a three-level medium. See Figs. 9-12. First, the addition of the third level leads to the disappearance of the SIT regime. The medium does not return to the ground state after the passage of the pulse due to the excitation of the third level. The question of preserving the SIT-like regime in a three-level medium for unipolar pulses, especially unusual shaped pulses, is rather non-trivial. Numerical calculations show that it is possible to preserve SIT in the case of, for example, Gaussian pulses or hyperbolic secant pulses in a three-level medium [51,52- Arkhipov et al. 2024; Arkhipov et al. 2024].

In the case of rectangular pulses, the situation is more complicated. This is apparently due to the presence of a flat top in the pulse rather than a Gaussian pulse in the form of a half-wave of the field. It should be noted that, to our knowledge, the question of the characteristics of the SIT-like pulse propagation regime in a multilayer medium in the case of unipolar pulses of unusual shape has not yet been studied in detail in the literature due to the unusual temporal shape of these fields.

Numerical calculations for a rectangular pulse did not even show a qualitative agreement in the DM shape in the case of a three-level medium with respect to the two-level model. After the collision, the shape of the induced DM is already completely different from that in the case of a two-level medium, see Figs. 9-12. This is due to the fact that pulses do not propagate in the SIT-like regime in a three-level medium. They change shape more rapidly during propagation than in a two-level medium. The more complex behavior of polarization in a three-level medium with the addition of a third level also contributes to the differences.

Thus, by considering an additional level of the medium, numerical calculations have shown significant differences in the behavior of DM in strong fields. It should be noted that the similarities in the behavior of population gratings in two- and three-level media observed in [50- Arkhipov et al. 2024], when the Gaussian pulses did not overlap, were apparently caused by the fact that the third level was far away from the first two levels. Such a situation can be realized in the case of deep quantum wells. Apparently, differences in the behavior of DM should be expected when considering a larger number of levels in atomic media with a converging level structure.

**Conclusions**

In this paper, for the first time to our knowledge, a comparison has been made of the dynamics of microcavities in a two- and three-level medium during the collision of unipolar pulses of an unusual shape (rectangular) in this medium. A qualitative agreement in the behavior of the DM in the weak field regime in a two-level and in a three-level medium is shown. In contrast to the case of Gaussian pulses, for which the introduction of the third level did not have such a sensitive effect on the shape of the DM, changes in the shape of dynamic resonators were detected in a three-level medium.

In strong fields with significant excitation of the medium, a significant difference was also found. In this case, the SIT-like regime of the propagation of a rectangular pulse, which is predicted by a simple analytical solution of the equations for a two-level medium, disappears when the third level is taken into account, as shown by the results of our calculations. This is also a peculiarity of the pulse shape, which is rectangular. It differs from the case of Gaussian and hyperbolic secant pulses, which we have analyzed previously [51,52- Arkhipov et al. 2024; Arkhipov et al. 2024].

The third level is populated. This radically changes the behavior of the polarization of the medium. If, in the case of a two-level medium, pulse collision did not lead to the formation of Bragg population gratings outside the overlapping region due to the SIT regime, then in the case of rectangular pulses such behavior occurred immediately after the first pulse collision. Complex atomic population Bragg gratings appeared outside the pulse overlap region after the second and subsequent pulse collisions.

The studies carried out raise the question of the characteristics of the SIT-like regime of propagating unipolar light pulses of unusual shape in multilevel media. The answer to this question will be the subject of future studies. The results discussed above open up new possibilities for using unusual shaped pulses in ultrafast optics, such as light trapping [64- Arkhipov et al. 2022], ultrafast optical switching [Hassan et al. 2024] and petahertz electronics [Heide et al. 2024]. The DMs under study are of interest for the creation of optical memory systems based on atomic coherence with unipolar pulses. This scheme may have more advantages than other schemes based on photon echo using long multi-cycle pulses [67-69-

Moiseev et al. 2024, Moiseev et al. 2024, Moiseev et al. 2024], due to the possibility of rapidly controlling the properties of the induced DMs using half-cycle pulses [46-Arkhipov et al. 2024].

**Funding**

The research was carried out with financial support from the Russian Science Foundation within the framework of scientific project 23-12-00012 (numerical simulations for three-level medium) and State Assignment of Ioffe Institute, topic 0040-2019-0017 (numerical simulations for two-level medium).

**Data availability**

No datasets were generated or analysed during the current study.

**Author Contributions**

All authors contributed to the study conception and design.

**References**

1. Midorikawa, K.: Progress on table-top isolated attosecond light sources, Nature Photonics. 16, 267 (2022).
2. Kim, H. Y., Garg, M., Mandal, S., Seiffert, L., Fennel, T., Goulielmakis, E.:, Attosecond field emission, Nature. 613, 662 (2023).
3. S. Severino, K. Ziems, M. Reduzzi, A. Summers, H.W. Sun, Y.-H. Chien, S. Gr̈afe, and J. Biegert, Attosecond core-level absorption spectroscopy reveals the electronic and nuclear dynamics of molecular ring opening, Nature Photonics. 18, 731 (2024).
4. L' Huillie, A.: Nobel Lecture: The route to attosecond pulses, Reviews of Modern Physics. 96, 030503 (2024).
5. Krausz, F.: Nobel lecture: Sub-atomic motions, Reviews of Modern Physics. 96, 030502 (2024).

6. Agostini, P.: Nobel Lecture: Genesis and applications of attosecond pulse trains, Reviews of Modern Physics. 96, 030501 (2024).

7. NobelPrize.org URL: https://www.nobelprize.org/prizes/physics/2023/press-release

8. Jackson, J.D.. Classical Electrodynamics (John Wiley & Sons, 1962)].

9. Bessonov ,E.G.: On a class of electromagnetic waves, Sov. Phys. JETP. 53, 433 (1981).

10. Rosanov, N.N.: Area of ultimately short light pulses Opt. Spectr. 107, 721 (2009).

11. Rosanov, N.N., Arkhipov, R.M., Arkhipov M.V.: On laws of conservation in the electrodynamics of continuous media (on the occasion of the 100th anniversary of the SI Vavilov State Optical Institute), Phys. Usp. 61, 1227 (2018)].

12. Ginzburg, V. L., Syrovatskii, S. I.: Cosmic magnetic bremsstrahlung (synchrotron radiation), Sov. Phys. Usp. 8 674–701 (1966).

13. Wu, H.-C., Meyer-ter Vehn J.: Giant half-cycle attosecond pulses. Nature Photon. 6, 304 (2012).

14. Hassan, M.T., Luu, T.T., Moulet, A., Raskazovskaya, O., Zhokhov, P., Garg M., Karpowicz, N., Zheltikov, A.M., Pervak, V., Krausz, F., Goulielmakis, E.: Optical attosecond pulses and tracking the nonlinear response of bound electrons, Nature. 530, 66 (2016).

15. Xu, J., Shen, B., Zhang, X., Shi, Y., Ji, L., Zhang, L., Xu, T., Wang, W., Zhao, X., Xu, Z.: Terawatt-scale optical half-cycle attosecond pulses. Sci. Rep. 8, 2669 (2018).

16. Bogatskaya, A.V., Volkova, E.A., Popov, A.M.: Three-dimensional modeling of intense unipolar THz pulses formation during their amplification in nonequilibrium extended Xe plasma channel, Phys. Rev. E. 105, 055203 (2022).

17. Ilyakov, I., Shishkin, B.V., Efimenko, E.S., Bodrov, S.B., Bakunov, M.I.: Experimental observation of optically generated unipolar electromagnetic precursors, Optics Express. 30, 14978 (2022).

18. Kuratov, A.S., Brantov, A.V., Kovalev, V.F., Bychenkov, V.Yu.: Powerful laser-produced quasi-half-cycle THz pulses, Phys. Rev. E. 106, 035201 (2022).

19. Sazonov, S. V., Ustinov, N. V., Propagation of few-cycle pulses in a nonlinear medium and an integrable generalization of the sine-Gordon equation, Phys. Rev. A. 98, 063803 (2018).

20. Arkhipov, M., Pakhomov, A., Arkhipov, R., Rosanov, N. : Generation of an ultrahigh-repetition-rate optical half-cycle pulse train in the nested quantum wells, Optics Letters. 48, 4637 (2023).

21. Glazov, M.M., Rosanov, N.N.: Generation of unipolar electromagnetic pulses in semiconductor nanocrystals, Phys. Rev. A. 109 (5), 053523 (2024).

22. Rosanov, N.N.: Formation of electromagnetic pulses with nonzero electrical area by media with ferromagnetism, Optics Lett. 49 (6), 1493 (2024).

23. Arkhipov, M. V., Tsypkin, A. N., Zhukova, M. O., Ismagilov, A. O., Pakhomov, A. V., Rosanov, N.N., Arkhipov, R. M.: Experimental determination of the unipolarity of pulsed terahertz radiation. JETP Letters. 115(1), 1-6 (2022).

24. Gorelov, S. D., Novokovskaya, A. L., Bodrov, S. B., Sarafanova, M. V., Bakunov, M. I.: Unipolar fields produced by ultrafast optical gating of terahertz pulses, Appl. Phys. Lett. 126, 011104 (2025).

25. Rosanov, N., Tumakov, D., Arkhipov, M., Arkhipov, R.: Criterion for the yield of micro-object ionization driven by few-and subcycle radiation pulses with nonzero electric area, Phys. Rev. A. 104 (6), 063101 (2021).

26. Pakhomov, A., Arkhipov M., Rosanov, N., Arkhipov, R.: Ultrafast control of vibrational states of polar molecules with subcycle unipolar pulses, Phys. Rev. A. 105, 043103 (2022).

27. Arkhipov, R., Belov, P., Pakhomov, A., Arkhipov, M., Rosanov, N.: JOSA B. 41 (1), 285 (2024).

28. Alexandrov, I.A., Rosanov, N.N.: Vacuum pair production in zeptosecond pulses: Peculiar momentum spectra and striking particle acceleration by bipolar pulses, Phys. Rev. D. 110, L111901 (2024).

29. Arkhipov, R.M., Arkhipov, M.V., Rosanov, N.N.. Unipolar light: existence, generation, propagation, and impact on microobjects. Quant. Electron. 50 (9), 801 (2020).

30. Arkhipov, R.M., Arkhipov, M.V., Pakhomov, A.V., Obraztsov, P.A., Rosanov, N.N.. Unipolar and subcycle extremely short pulses: recent results and prospects (brief review), JETP Letters. 117 (1), 8 (2023).

31. Rosanov, N.N.. Unipolar pulse of an electromagnetic field with uniform motion of a charge in a vacuum, Phys. Usp. 66, 1059 (2023).

32. Rosanov, N.N., Arkhipov, M.V., Arkhipov, R.M., Pakhomov, A.V. Half-cycle electromagnetic pulses and pulse electric area, Contemprorary Physics. **64** (3), 224 (2023).

33. Rosanov, N.N., Arkhipov, M.V., Arkhipov, R.M.: Extremely short and unipolar light pulses: state of the art, Phys. Usp. 67 (11) 1129 - 1138 (2024).

34. Rosanov, N. N., Arkhipov, M. V., Arkhipov, R. M.: Terahertz Photonics, edited by V. Ya. Panchenko, A. P. Shkurinov (RAS, Moscow, 2023), pp. 360–393 (in Russian).

35. Pakhomov, A., Arkhipov, M., Rosanov, N., Arkhipov R.: Generation of waveform-tunable unipolar pulses in a nonlinear resonant medium, Physical Review A. 106, 053506 (2022).

36. Arkhipov, R. M., Arkhipov, M. V., Pakhomov, A. V., Diachkova, O., Rosanov N. N.: Radiation of a solitary polarization pulse moving at the speed of light, JETP Letters. 117, 574 (2023).
37. Pakhomov, A., Rosanov, N., Arkhipov, M., Arkhipov, R.: Sub-10 fs unipolar pulses of a tailored waveshape from a multilevel resonant medium, Optics Letters. 48, 6504 (2023).
38. Arkhipov, R., Pakhomov, A., Arkhipov, M., Demircan, A., Morgner, U., Rosanov, N., Babushkin, I.: Selective ultrafast control of multi-level quantum systems by subcycle and unipolar pulses. Optics Express. 28 (11), 17020 (2022).
39. Bastrakova, M.V., Klenov, N.V., Satanin, A.M.. One-and two-qubit gates: Rabi technique and single unipolar pulses, Phys. Solid State, 61, 1515 (2019).
40. Bastrakova, M.V., Klenov, N.V., Satanin, A.M.. Tomography of qubit states and implementation of quantum algorithms by unipolar pulses, JETP. 131, 507 (2020).
41. Arkhipov, R.M., Arkhipov, M.V., Pakhomov, A.V., Dyachkova, O.O., Rosanov, N.N.: Nonharmonic Spatial Population Difference Structures Created by Unipolar Rectangular Pulses in a Resonant Medium, Opt. Spectr. 130 (11), 1443 (2022).
42. Diachkova, O.O., Arkhipov, R.M., Arkhipov, M.V., Pakhomov, A.V.,. Rosanov, N.N.: Light-induced dynamic microcavities created in a resonant medium by collision of non-harmonic rectangular 1-fs light pulses. Opt. Commun. 538, 129475 (2023)
43. Diachkova, O., Arkhipov, R., Pakhomov, A., Rosanov, N.: Optical microcavity formation and ultrafast control using half-cycle attosecond pulses in two-and three-level media. Opt. Commun. 565, 130666 (2024).
44. Arkhipov, R., Pakhomov, A., Diachkova, O., Arkhipov, M., Rosanov, N.: Bragg-like microcavity formed by collision of single-cycle self-

induced transparency light pulses in a resonant medium. Opt. Lett. 49 (10), 2549 (2024).

45. Arkhipov, R.M.: Dynamics of Atomic Population Gratings in Collisions of Unipolar Light Pulses in a Multilevel Resonant Medium, Bulletin of the Lebedev Physics Institute, 51 (5), S366 (2024);

46. Arkhipov, R., Pakhomov, A., Diachkova, O., Arkhipov, M., Rosanov, N.: Analytical and numerical study of light-induced optical microcavity generation by half-cycle light pulses in the resonant medium, JOSA B, 41 (8), 1721 (2024).

47. Arkhipov, R.M., Dyachkova, O.O., Arkhipov, M.V., Pakhomov, A.V., Rosanov, N.N., Optical microresonators created by unipolar light pulses in a medium (review), Opt. Spectr., 132 (9), 919 (2024) (in Russian); Arkhipov, R. M., Arkhipov, M. V., Pakhomov, A. V., Diachkova O. O., Rosanov, N. N., Interference of the Electric and Envelope Areas of Ultrashort Light Pulses in Quantum Systems, Radiophys Quantum El. 66, 286–303 (2023). https://doi.org/10.1007/s11141-024-10295-x

48. Arkhipov, R.M., Arkhipov, M.V., Rosanov, N.N.: The influence of the polarity of half-cycle pulses on the dynamics of microresonators in a three-level medium, Opt. Spectr., 132 (9), 938 (2024) (in Russian).

49. Arkhipov, R., Arkhipov, M., Pakhomov, A., Diachkova, O., Rosanov, N.: Generation and control of population difference gratings in a three-level hydrogen atomic medium using half-cycle attosecond pulses nonoverlapping in the medium. Phys. Rev. A. 109, 063113 (2024).

50. Arkhipov, R.M., Arkhipov, M.V., Rosanov, N.N.: Comparison of the dynamics of population difference gratings created in a two-level and three-level medium by half-cycle light pulses, Opt.Spectr. 132(4), 399 (2024).

51. Arkhipov, R. M., Pakhomov, A. V., Arkhipov, M. V., Rosanov, N. N.: Coherent propagation of half-cycle light pulse in a three-level medium JETP., 166(8), 174 (2024) (in Russian).

52. Arkhipov, R. M., Arkhipov, M.V., Rosanov, N.N.: Opt. Spectr., 132(11), 1196 (2024) (in Russian).

53. Yariv A.. Quantum Electronics (John Wiley & Sons, N.Y., London, Toronto, 1975).

54. Bayer, M. Forchel, A., Temperature dependence of the exciton homogeneous linewidth in In 0.60 Ga 0.40 As/GaAs self-assembled quantum dots, Phys. Rev. B. 65, 041308 (2002).

55. Babbit, W. R., Mossberg, T.: Time-domain frequency-selective optical data storage in a solid-state material Opt. Commun. 65, 185 (1988).

56. Frisch, S.E.. Optical spectra of atoms (State Publishing House of Physical and Mathematical Literature, Moscow-Leningrad, 1963) (in Russian).

57. Arkhipov, R. M., Rosanov, N. N., Creation of a dynamic microresonator by collision of half-cycle light pulses in a resonant medium, Optics and Spectroscopy, 128(5), 630-634 (2020) (in Russian).

58. Allen, L., Eberly, J.H.. Optical resonance and two-level atoms (Wiley, N.Y., 1975).

59. McCall, S. L., Hahn, E. L., Self-induced transparency. Physical Review, 183(2), 457 (1969).

60. Ahokas, J., Vainio, O., Järvinen, J., Khmelenko, V. V., Lee, D. M., Vasiliev, S.. Stabilization of high-density atomic hydrogen in $H_2$ films at T< 0.5 K, Phys. Rev. B. 79, 220505(R) (2009).

61. Järvinen, J., Khmelenko, V. V., Lee, D. M., Ahokas, J., Vasiliev, S.: Atomic Hydrogen in Thick H2 Films at Temperatures 0.05–2 K. Journal of Low Temperature Physics. 162, 96-104 (2011).

62. Arkhipov, R.M., Arkhipov, M.V., Babushkin, I., Demircan, A., Morgner, U., Rosanov, N.N.. Utrafast creation and control of population density gratings via ultraslow polarization waves. Opt. Lett., 41, 4983 (2016).

63. Arkhipov, R.M.: Electromagnetically induced gratings created by few-cycle light pulses (brief review). JETP Lett. 113, 611 (2021).

64. Arkhipov, M., Arkhipov, R., Babushkin, I., Rosanov, N.: Self-stopping of light. Phys. Rev. Lett., **128** (20), 203901 (2022).

65. Hassan, M.T.. Lightwave Electronics: Attosecond Optical Switching, ACS Photonics. 11, 334 (2024).

66. Heide, C., Keathley, P. D., Kling, M. F., Petahertz electronics. Nat.Rev. Phys. 6, 648 (2024).

67. Moiseev, S.A., Gerasimov, K.I., Minnegaliev, M.M., Moiseev, E.S., Deev, A.D., Balega, Yu.Yu.: Echo protocols of an optical quantum memory, arXiv preprint arXiv:2410.01664 (2024).

68. Moiseev, S.A., Gerasimov, K.I., Minnegaliev, M.M., Moiseev, E.S.: Optical quantum memory on macroscopic coherence, arXiv preprint arXiv:2408.09991 (2024).

69. Moiseev, S.A., Minnegaliev, M.M., Gerasimov, K.I., Moiseev, E.S., Deev, A.D., Balega, Yu.Yu.: Optical quantum memory on atomic ensembles: physical principles, experiments and possibilities of application in a quantum repeate, Phys. Usp., 67 (2024) DOI: 10.3367/UFNe.2024.06.039694].